# A Novel Image Steganographic Approach for Hiding Text in Color Images using HSI Color Model

*Khan Muhammad, Jamil Ahmad, Haleem Farman, Muhammad Zubair
Department of Computer Science, Islamia College Peshawar, Pakistan
Corresponding Author email: *khan.muhammad.icp@gmail.com

**ABSTRACT**
Image Steganography is the process of embedding text in images such that its existence cannot be detected by Human Visual System (HVS) and is known only to sender and receiver. This paper presents a novel approach for image steganography using Hue-Saturation-Intensity (HSI) color space based on Least Significant Bit (LSB). The proposed method transforms the image from RGB color space to Hue-Saturation-Intensity (HSI) color space and then embeds secret data inside the Intensity Plane (I-Plane) and transforms it back to RGB color model after embedding. The said technique is evaluated by both subjective and Objective Analysis. Experimentally it is found that the proposed method have larger Peak Signal-to Noise Ratio (PSNR) values, good imperceptibility and multiple security levels which shows its superiority as compared to several existing methods.

## 1. INTRODUCTION

The word steganography is derived from two Greek words; "stegano" meaning protected and "graphia" meaning writing. It can be defined as the process of writing messages in a way in which the presence of secret message is known only to sender and receiver. Steganography require a carrier object, secret data and embedding algorithm. It may also need an encryption algorithm and secret key in some cases in order to increase the security levels of steganography. Applications of steganography includes secure transmission of top secret documents between national and international governments, securing online banking and voting systems, secret communication between criminals and terrorists and sending Trojan horses and viruses to attack on systems etc.[1-7].

### 1.1 Steganography VS Cryptography
Steganography and Cryptography both techniques are used for data confidentiality (protection of information from unwanted parties).However there also exists some differences between them that is described below.

i. Cryptography is the practice and study of secure communication but Steganography is an art as well as a science of covert communication.

ii. The main focus of Cryptography is to keep the contents of the data secret while Steganography aims to keep the existence of the data secret.

iii. Security in cryptography is achieved by converting the secret data into non-understandable form while in steganography; security is achieved by hiding data in apparently harmless carriers to hide the existence of data.

iv. Cryptography provides us secure communication by using a key to read the information. An intruder is unable to remove the used encryption but he/she can easily modify the encrypted file and can make the encrypted file meaningless which cannot be easily understood by the receiver. On the other hand, Steganography provide us a method of covert communication that can be removed only if the carrier object in which data has been embedded, is modified. The confidentiality of hidden data will remain as it is, until and unless a suspicious user is succeeded in finding a method for detecting it.

v. The final endeavor of cryptography is a cipher text but the final output of steganography is a stego object.

vi. To break the cryptography, we compare some sections of the plaintext with the sections of the cipher text but in breaking steganography we compare the cover object with the stego object plus some possible sections of the message.

vii. Cryptography fails when an intruder gain access to the contents of the cipher material but steganography fails only when a malicious user detects the presence of the secret data. The Table 1 given below also clarifies the difference between these techniques **[8-10]**.

Table 1: Comparisons of Cryptography and Steganography

| Comparison Table | | | |
|---|---|---|---|
| **Method/Property** | Confidentiality | Integrity | Unremovability |
| **Cryptography** | Yes | No | Yes |
| **Steganography** | Yes/No | Yes/No | Yes |

**Note:** *None of the above methods alone can be perfect and compromised.*

## 1.2 Types of Steganography w.r.t carrier object

Steganography embeds secret data into digital carriers like image, audio, video etc such that it cannot be easily detected by the Human Visual System (HVS). There are five types of steganography on the basis of carrier object that is used for embedding the secret data. These types are briefly described and its diagrammatic representation is given in Fig.1.

### 1.2.1 Image Steganography
The type of Steganography in which image is used as cover object is called Image Steganography. Generally, the techniques in this method modify the image pixels for hiding secret information. Images are considered to be the best cover objects/carriers for hiding information because it contains large amount of redundant bits.

### 1.2.2 Network Steganography
The type of Steganography in which network protocol (TCP, IP, UDP and ICMP etc) is used as cover object is called Network Steganography. In this method, information is hidden in some fields of the Header of TCP/IP packet that are optional or never used.

### 1.2.3 Text Steganography
Text Steganography uses text as a carrier for hiding secret information. In this technique secret message is hidden in the $n^{th}$ letter of every word of carrier text. Unlike other types of steganography (audio, video, network and image), Text steganography is much more difficult because of less redundancy in the text. Text steganography is preferable for simple communication where less memory is needed.

### 1.2.4 Audio Steganography
The type of steganography in which audio is used for encoding and decoding of secret data is called audio steganography. This type of steganography is considered as a significant medium for secret communication due to attractiveness of voice IP (VOIP). The different types of audio formats used for audio steganography are MPEG, MIDI, WAVE, AVI, etc.

### 1.2.5 Video Steganography
When video is used as a carrier for information hiding, then the steganography used is called video steganography. This approach is capable to hide large volume of data as video is the combination of frames/images and contains large amount of redundant bits. The formats used by video steganography are H.264, Mp4, MPEG, AVI etc.

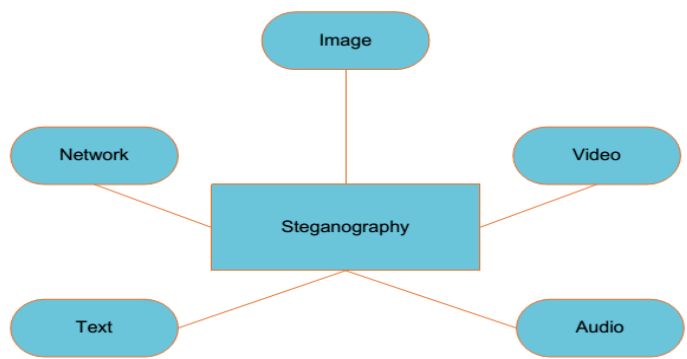

**Figure 1:** Types of Steganography w.r.t carrier object

This paper proposes a novel method for steganography to overcome the limitations of existing steganographic methods. Image has been chosen as a carrier object in this paper because it contains more redundant bits. The rest of the paper is organized as follows. Section 2 critically discusses some existing steganographic methods in literature whose defects led us towards current proposed work. The proposed technique is detailed in section 3. Section 4 is devoted to experimental results and discussion. The conclusion of the paper is given at the end in section 5.

## 2. LITERATURE REVIEW

The simplest method to hide secret data inside a cover image is LSB. In this method the least significant bits of the carrier image pixels are replaced with the secret data bits. Payload capacity of LSB method can be increased if more than 1 LSBs are used for message embedding but it brings noticeable changes in the carrier image. LSB method is simple to implement but it is vulnerable to many statistical attacks like RS, image processing operations and Chi-Square analysis etc.[11-17].

Adnan Abdul-Aziz Gutub proposed a more robust steganographic technique in [18] in which one channel is used for indication while other two channels are used for embedding secret data in a predefined cycle manner which enhances the robustness of proposed method. The experimental results show us the high payload capacity and better imperceptibility of the proposed algorithm. This method also avoids the key exchange overhead.

In [19], the authors propose a robust method which embeds variable bits in image pixels depending on the pixel value and value of mean and standard deviation (SD). Two bits are inserted in the image pixel if (mean−SD/2) is greater than pixel value; 3 bits are stored if (mean +SD/2) is greater than pixel value otherwise 4 bits are stored in each pixel value. Chaotic effect is also obtained in the proposed method by using random traversing path which make the attack loathsome but nothing is given about generating the random traversing path.

Grover et al., in [20], presents an adaptive edge based LSB substitution technique which hides three(3) bits of secret data in edgy pixel and two(2) secret bits in non-edgy pixels in blue channel of the RGB image. This method has high payload capacity and is more robust as compared to simple LSB substation method because secret data is divided into two sets first and then it is embedded in cover image starting from the central pixel and traversing through the whole image which increases its robustness.

In [21], the authors proposed a new method to embed secret data in the GREEN or BLUE channel of carrier image on the basis of secret key bits and RED channel LSB. This method adds one more level security to the existing LSB method by utilization of secret key. The RED channel LSB and secret key bit is xored and then a decision is taken on the basis of its result to replace the LSB of GREEN or BLUE channel. The proposed method has the same payload, more robustness and better security as compared to simple LSB method. However the secure key exchange of secret key is an open challenge and is an extra overhead of proposed method.

Ibrahim and Kuan have developed a SIS (Steganography Imaging System) in [22] which uses a secret key to enhance the security of the proposed system. The authors have made use of zipping to zip the secret key and secret data in order to increase the payload. The zip file is then converted into bits stream and hidden in cover image. The proposed algorithm has high payload capacity and better quality of stego images but this technique is proposed only for BMP format images.

Thanikaiselvan and Arulmozhivarman have proposed a more robust approach for color images in transform domain based on Reversible Integer Haar Wavelet Transform (RIHWT) and Graph Theory (GT)[23]. The proposed method applies RIHWT on RED, GREEN and BLUE channels separately and selects wavelets coefficients based on Graceful Graph (GG) for randomly scattering the secret data in wavelets coefficients. Three different unique keys are used for encoding and decoding of secret data in the proposed method. Key1 is subband selection key for selecting one of the sub bands i.e. LL, LH, HL and HH; Key2 is for selecting coefficients inside the selected subband randomly on the basis of GG; Key3 is for deciding the number of bits to be stored in selected coefficients. The proposed method gives us promising results in terms of imperceptibility, robustness and payload.

## 3. PROPOSED ALGORITHM

All the communicating bodies want the confidentiality, integrity and authenticity of their secret information. Different approaches are used to cope with these security issues like digital certificate, digital signature and cryptography. But these methods alone cannot be compromised. Steganography is the best solution to these problems as it hides the existence of secret data. This paper proposes a novel image steganography LSB based technique for RGB images using color space exchanging from RGB to HSI. The secret data is embedded in I-Plane of HSI color model using LSB method. Finally the resultant image is retransformed to RGB color model to make the stego image.

### 3.1 Color Models

Color models are also known as color systems or color spaces. The main goal of these color spaces is to represent all colors in a standard way. A color model is a way of representing a set of colors mathematically. The most popular color spaces are RGB, YCbCr (Luminance Component, Chroma Blue difference, Chroma Red difference) and CMYK (Cyan, Magenta, Yellow and Black), HSI (Hue, Saturation and Intensity).

The HSI color model is derived from RGB color space that represents colors the way the human eyes perceive and interpret colors. Human eye describes colors by its hue, saturation and intensity. Hue represents a pure color i.e. pure red, yellow etc. Saturation gives us a measure of the degree to which a pure colour is diluted by white light. Intensity is the brightness of a color[24, 25].

#### 3.1.1 Why HSI Color Model for Embedding

The proposed method uses HSI color space for information hiding because of the following reasons.
- ➢ Processing an image in RGB color system is relatively more difficult and time consuming. All the three values of a particular pixel need to be read, the intensity is then calculated, the desired changes is made, new RGB values are recalculated and stored.
- ➢ The brightness information in RGB color space is embedded in its each layer which indicates that all the three layers are strongly correlated to one another and any changes to one of its layer will have its corresponding effect on other layers.
- ➢ To generate a particular color in RGB colour space, all the three components must be of equal bandwidths.
- ➢ RGB is not an efficient color space when we are concerned with real-world images.
- ➢ To simply processing, programming and end user manipulations.

#### 3.1.2 Conversion from RGB to HSI

The image in RGB color space is transformed into HSI color space by the following formulae.

$$r = \frac{R}{R+G+B} \quad (1) \qquad g = \frac{G}{R+G+B} \quad (2) \qquad b = \frac{B}{R+G+B} \quad (3)$$

$$h = \cos^{-1}\left[\frac{0.5 \times \{(r-g)+(r-b)\}}{[(r-g)^2+(r-b)(g-b)]^{\frac{1}{2}}}\right] \quad h \in [0,\pi] \quad for\ b \leq g \quad (4)$$

$$h = 2\pi - \cos^{-1}\left[\frac{0.5 \times \{(r-g)+(r-b)\}}{[(r-g)^2+(r-b)(g-b)]^{\frac{1}{2}}}\right] \quad h \in [\pi, 2\pi] \quad for\ b > g \quad (5)$$

$$s = 1 - 3 \times \min(r,g,b) \quad s \in [0,1] \quad (6)$$

$$i = \frac{R+G+B}{3 \times 255} \quad i \in [0,1] \quad (7)$$

For the sake of convenience h, s and i values are transformed into these ranges [0,360], [0,100], [0, 255], respectively by:

$$H = \frac{h \times 180}{\pi} \quad (8) \qquad S = s \times 100 \quad (9) \qquad H = h \times 255 \quad (10)$$

#### 3.1.3 Conversion from HSI to RGB

$$h = \frac{H \times \pi}{180} \quad (11) \qquad s = \frac{S}{100} \quad (12) \qquad i = \frac{I}{255} \quad (13)$$

$$x = i \times (1 - s) \quad (14) \qquad y = i \times \left[1 + \frac{s \times \cos(h)}{\cos\left(\frac{\pi}{3} - h\right)}\right] \quad (15) \qquad z = 3i - (x + y) \quad (16)$$

If $h < \frac{2\pi}{3}$ then $\quad b = x;\ r = y \quad$ and $g = z$

If $\frac{2\pi}{3} \leq h < \frac{4\pi}{3}$ then $\quad h = h - \frac{2\pi}{3} \quad r = x,\ g = y$ and $\quad b = z$

These are the normalized values of r, g and b in the range [0, 1]. Finally these values are multiplied by 255 in order to form the original RGB values[24].

### 3.1.4 Embedding Algorithm
*Input:* Cover Colour Image, Secret data
*Output:* Stego Image
Step 1: Take the cover RGB image and secret data.
Step 2: Convert the RGB image into HSI color model using section 3.1.2 formulas.
Step 3: Convert the secret data into 1-D array of bits.
Step 4: Take a pixel from I-Plane and replace its LSB with a secret bit.
Step 5: Repeat Step 4 until and unless all secret bits are encoded in the I-Plane pixels.
Step 6: Convert the HSI image into RGB color space using the formulae of section 3.1.3.
Step 7: Write the stego image.

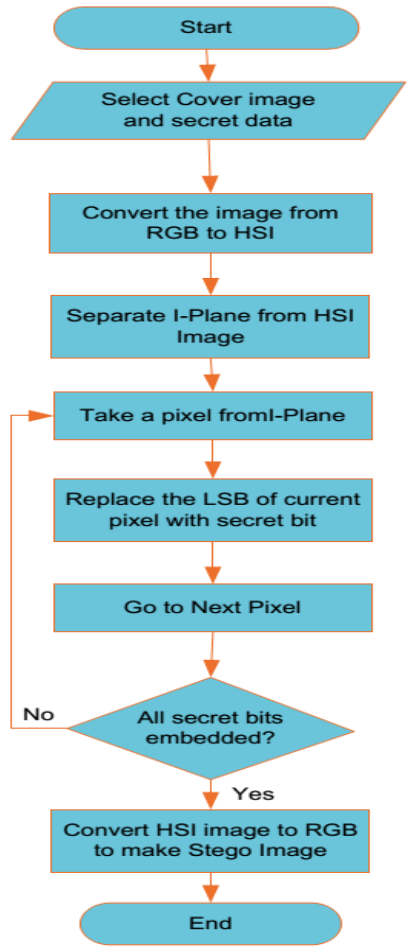

**Figure 2:** Embedding Algorithm Flowchart

### 3.1.5 Extraction Algorithm

*Input:* Stego Image
*Output:* Secret data
Step 1: Take the stego image and convert it into HSI color space.
Step 2: Consider the I-Plane only for extraction of secret data.
Step 3: Extract the LSB of current pixel from I-Plane of HSI image.
Step 4: Repeat Step 3 until and unless all secret bits are decoded.
Step 5: Convert secret bits into secret data i.e. text, image etc.

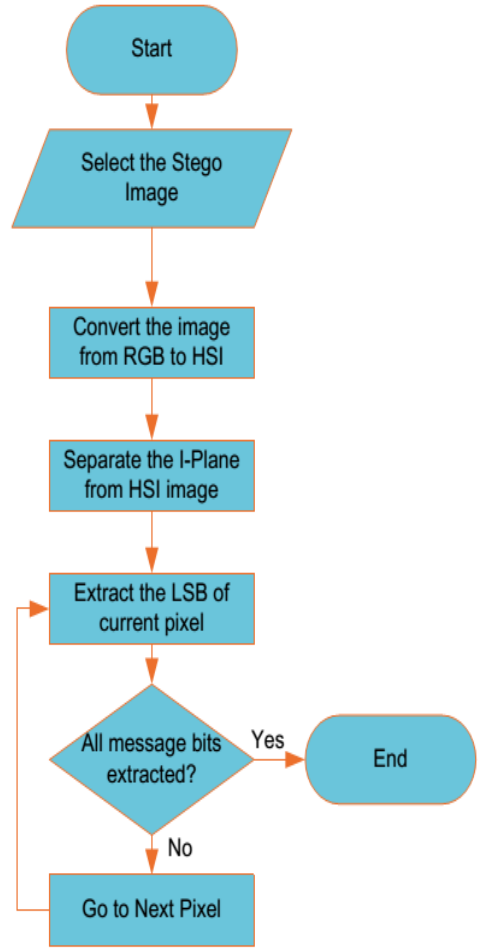

**Figure 3:** Extraction Algorithm Flowchart

## 4. EXPERIMENTAL RESULTS AND DISCUSSION

The proposed method, LSB technique and technique in [21] are simulated using MATLAB R2013a. For experiments we have embedded variable amount of cipher in different standard color images of same and different dimensions to estimate the performance of the proposed technique. The proposed technique is evaluated by 3 different perspectives; hiding the same amount of cipher in different images of the same dimensions; hiding variable amount of cipher in the same image of the same dimension and hiding same amount of cipher in the same image of different dimensions. The standard color images used for experiments are lena.png, baboon.png, peppers.png, trees.tiff etc.

### 4.1 Comparison of proposed method with existing methods

The comparison among proposed algorithm, simple LSB and algorithm in [21] is based on two types of analysis named as subjective analysis and objective analysis. Subjective analysis is done using Human Visual System (HVS) to notice the changes between the cover and stego images and their corresponding histograms. A few samples of standard color cover and stego images and their histograms for the proposed method are shown below in Figure 9, Figure 10 and

Figure 11. From figures it is observed that there is no noticeable change in the cover and stego images and their histograms which shows the effectiveness of the proposed method.

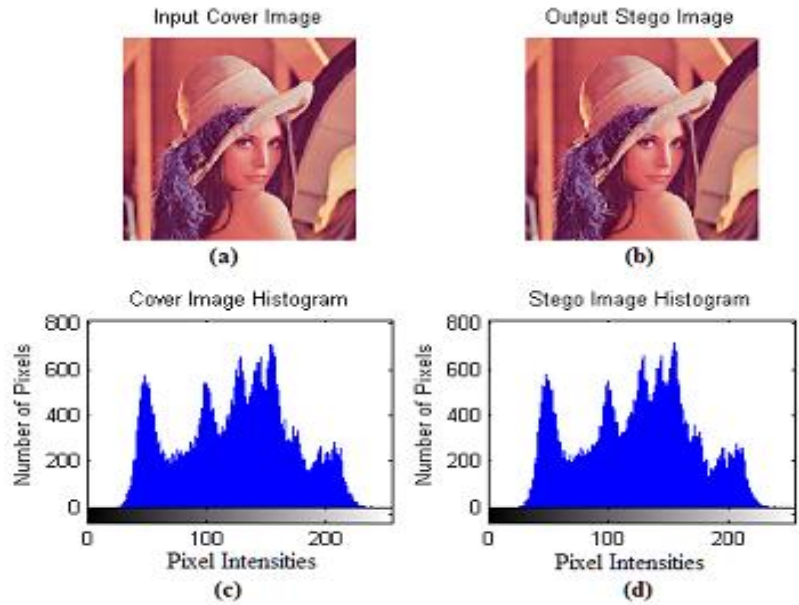

**Figure 9:** Lena cover and stego image and their histograms

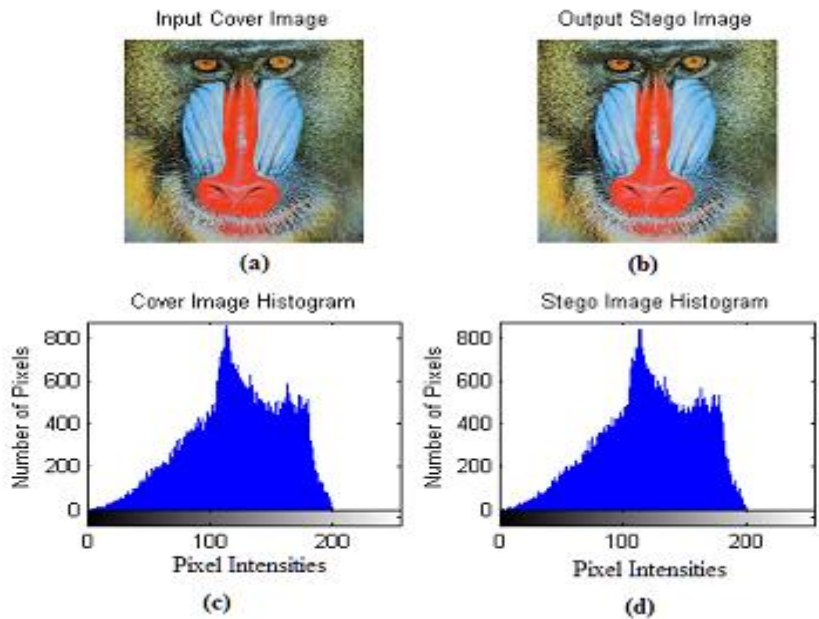

**Figure 10**: baboon cover and stego image and their histograms

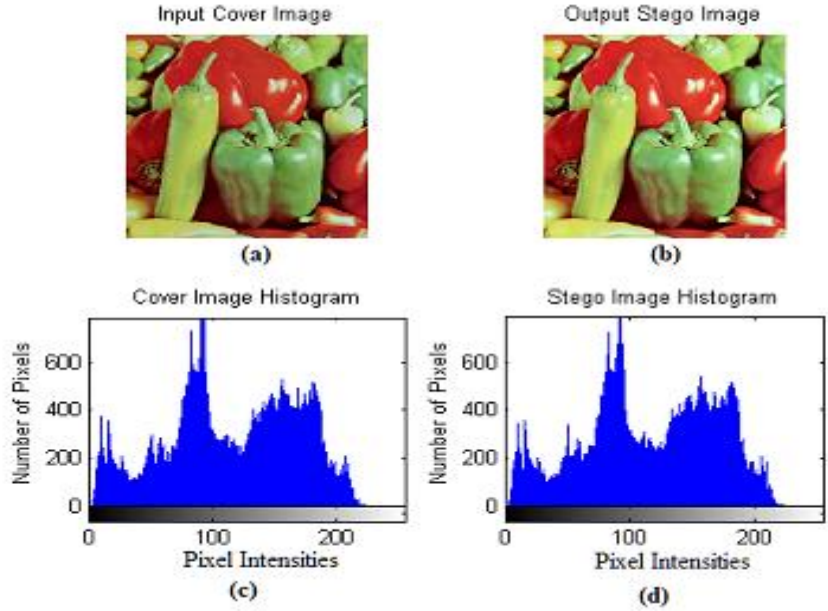

**Figure 11:** Peppers cover and stego image and their histograms

Objective analysis is a mathematical standard for measuring the distortion that occurs in the cover image after embedding secret data[26]. Objective analysis is performed on the proposed method using Peak Signal-to-Noise Ratio (PSNR) and Mean Square Error (MSE). The PSNR and MSE are calculated by the following formulas of (1) and (2).

$$PSNR = 10 \log_{10}\left(\frac{C_{\max}{}^2}{MSE}\right) \qquad (1)$$

$$MSE = \frac{1}{MN} \sum_{x=1}^{M} \sum_{y=1}^{N} (S_{xy} - C_{xy}) \qquad (2)$$

Note that here $M$ and $N$ are image dimensions, $x$ and $y$ are loop variables, $S$ is stego image, $C$ is cover image and $C_{max}$ is the maximum pixel intensity among both images [27-33]. The experimental results of the proposed methods, LSB and method in [21] are shown in Table 2, Table 3 and Table 4 respectively.

**Table 2:** Comparison of proposed method with LSB and Method in [21] based on PSNR

| Image Name | LSB Method | Karim's Method[21] | Proposed Method |
|---|---|---|---|
| | PSNR (dB) | PSNR (dB) | PSNR (dB) |
| baboon.png | 61.8784 | 48.558 | 94.4421 |
| lena.png | 42.6331 | 42.6204 | 42.5417 |
| peppers.png | 62.9666 | 17.39 | 100 |
| building.png | 51.4677 | 47.0305 | 70.0942 |
| parrot.png | 49.708 | 49.8421 | 62.4971 |
| trees.tiff | 63.46 | 50.5301 | 83.9069 |

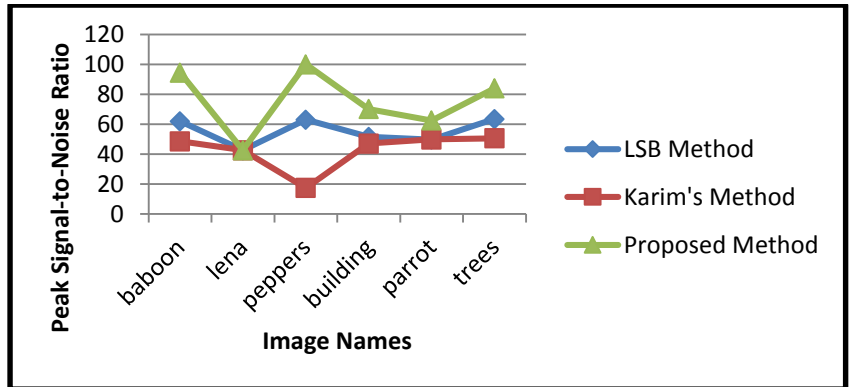

**Figure 12:** Comparison based on PSNR with different images of dimension 256×256

**Table 3:** Comparison based on PSNR with variable amount of cipher embedded

| Image Name | Cipher size in (KBs) | LSB Method PSNR(dB) | Karim's Method [21] PSNR(dB) | Proposed Method PSNR (dB) |
|---|---|---|---|---|
| baboon with dimension 256×256 | 2 | 63.3775 | 52.0373 | 83.6503 |
|  | 4 | 61.8442 | 51.6345 | 78.4215 |
|  | 6 | 60.4909 | 51.1776 | 74.3139 |
|  | 8 | 59.7481 | 50.8811 | 73.6444 |

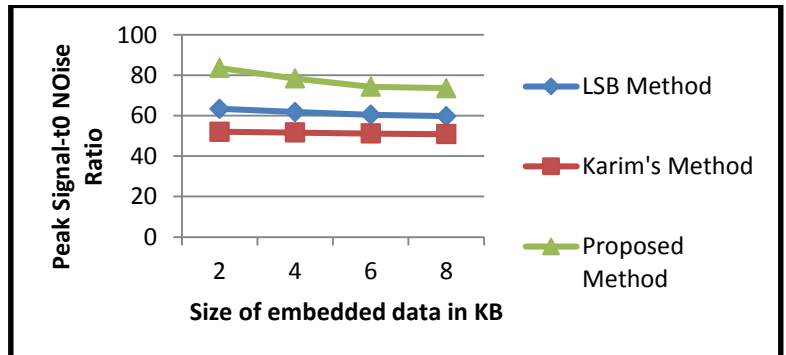

**Figure 13:** Comparison based on PSNR with same image dimension and variable amount of cipher

**Table 4:** PSNR based comparison with same size cipher and different image dimensions

| Image Dimensions | LSB Method PSNR(dB) | Karim's Method [21] PSNR(dB) | Proposed Method PSNR (dB) |
|---|---|---|---|
| 128×128 | 70.3187 | 65.5328 | 47.1518 |
| 256×256 | 61.8784 | 50.8811 | 73.6444 |
| 512×512 | 52.4555 | 37.2456 | 86.3986 |
| 1024×1024 | 59.204 | 41.9577 | 65.2814 |

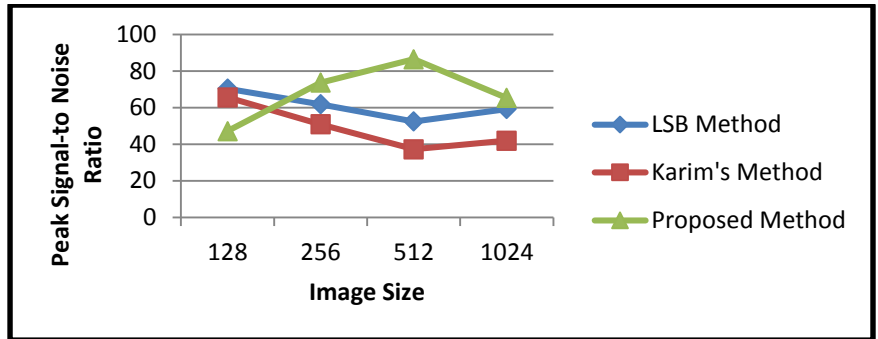

**Figure 14:** PSNR based comparison with same size cipher different image dimensions

## 5. CONCLUSION

This paper proposed a novel approach of image steganography for true color images with better imperceptibility, security and robustness. The said approach uses the HSI color model to hide secret messages inside color images to increase the security of the proposed technique. An average PSNR of 75.57dB is achieved with this novel approach which shows the superioty of the proposed method as compared to existing methods. This method introduces and adds an extra security level barrier in the way of an attacker which makes the attack on this algorithm awful and misguides the process of steganalysis.

## 6. ACKNOWLEGMENT

The authors wish to thank all the contributors for their critical and technical review of the proposed work and their valuable support and guidance.